\documentclass[a4paper, 11pt]{article}
\pdfoutput=1

\usepackage{tikz}
\usetikzlibrary{calc}

\usepackage[compat=1.1.0]{tikz-feynman}

\usepackage{jheppub}

\usepackage{verbatim}
\usepackage{amsmath,amsthm,amssymb}
\usepackage{natbib}
\usepackage{physics}
\usepackage{cancel}
\usepackage{amsmath,amsthm,amssymb}
\usepackage[nottoc]{tocbibind}
\usepackage{bm}
\usepackage{graphicx}
\usepackage[font=small,labelfont=bf]{caption}
\usepackage{hyperref}
\hypersetup{
    colorlinks,
    citecolor=blue,
    filecolor=blue,
    linkcolor=blue,
    urlcolor=blue
}

\def\beq{\begin{equation}}

\def\eeq{\end{equation}}

\begin{document}

\title{On running coupling in the JIMWLK evolution and its Langevin formulation}

\author[a]{Tolga Altinoluk,}
\affiliation[a]{National Centre for Nuclear Research, 02-093 Warsaw, Poland}

\author[a]{Guillaume Beuf,}

\author[b]{Michael Lublinsky,}
\affiliation[b]{Physics Department, Ben-Gurion University of the Negev, Beer Sheva 84105, Israel}

\author[c,d]{Vladimir~V.~Skokov}

\affiliation[c]{Department of Physics, North Carolina State University, Raleigh, NC 27695, USA}

\abstract{
Various {
conventional running coupling prescriptions  reproducing $\beta_ 0$-dependent terms of NLO JIMWLK are reviewed and found to be theoretically inconsistent: the JIMWLK evolution Hamiltonian with running coupling violates  }
 the requirement of positive semidefiniteness. This requirement appears 
to be tightly related to the possibility of having a Langevin formulation for the evolution. 
{
We 
also review the scheme that attributes a part of $\beta_0$-dependent terms to the 
DGLAP evolution of the projectile. The remaining $\beta_0$-dependent contributions sum up into so-called the ``daughter dipole'' prescription, which leads to a manifestly positive semidefinite Hamiltonian.}
}   

\maketitle
\section{Introduction}

High energy evolution of a hadronic observable $\cal O$ in a dilute-on-dense scattering (DIS-like processes) is governed by the  JIMWLK   equation of the form 
\begin{equation}
\label{Eq:JIMWLKonO}
\dot{\cal O} \,=\,-\,H^{\rm  JIMWLK}\,{\cal O}\,,
\end{equation}
which is 
a non-linear functional generalization of  the BFKL  equation 
\cite{Lipatov:1985uk,Gribov:1983ivg,Balitsky:1978ic}. 

The JIMWLK evolution
\cite{Jalilian-Marian:1997qno,Jalilian-Marian:1997ubg,Kovner:2000pt,Iancu:2000hn,Ferreiro:2001qy}, 
or mainly its large $N_c$/mean field version, the Balitsky-Kovchegov (BK)
equation \cite{Balitsky:1995ub,Balitsky:1997mk,Kovchegov:1999yj,Balitsky:2007feb},
was instrumental in the phenomenology of high-energy QCD. Relaxing the large $N_c$/mean field approximation turns out to be very challenging. 
One is then forced to simulate the entire JIMWLK Hamiltonian, meaning simulating a 2+1 dimensional non-local QFT. 
So far, the stochastic formulation of the Leading Order (LO) JIMWLK in terms of Langevin equation  \cite{Blaizot:2002np,Rummukainen:2003ns}  is  the only  known 
approach to this computational problem. The approach is   valid only for fixed coupling constant or for very specific running coupling prescriptions \cite{Lappi:2011ju,Lappi:2012vw}. This is a severe limitation for any reliable phenomenology. 
Due to this issue, among others, most phenomenological studies involving the JIMWLK equation have so far been restricted to LO accuracy. Yet, in order to 
reveal gluon saturation effects, it is critical to improve the precision of the theoretical calculations
by 
including higher order contributions.
A lot of progress has been achieved in this direction 
over the last fifteen years. Particularly, the NLO formulation of both the BK equation 
\cite{Balitsky:2007feb} 
and the JIMWLK Hamiltonian \cite{Kovner:2013ona,Lublinsky:2016meo}  (including the contribution due to quark masses \cite{Dai:2022imf}) became available,
enabling the start of the era of NLO phenomenology
\cite{Lappi:2015fma,Lappi:2016fmu,Beuf:2020dxl,Hanninen:2022gje} based on the BK equation.
However, beyond LO, the JIMWLK Hamiltonian does not seem to admit any simple stochastic formulation.  
Consequently, there has been no phenomenological applications of the full NLO JIMWLK Hamiltonian.

 Despite the theoretical progresses, bringing all the calculations relevant to DIS systematically to the NLO accuracy remains a major challenge.
It was discovered that at NLO, there emerge  conceptual problems related to
 large energy-independent new logarithmic corrections, threatening  stability of the $\alpha_s$ expansion, and consequently requiring additional resummations.
 These logarithms could be divided into three groups. 
 The first one contains 
 large logs related to time ordering between gluon emissions. 
 A lot of efforts have been invested over the years in attempting  to impose time ordering via various kinematical constrains at the level of the BFKL, BK and even JIMWLK equations~\cite{Salam:1999cn,Kutak:2004ym,Motyka:2009gi,Beuf:2014uia,Iancu:2015vea,Iancu:2015joa,Hatta:2016ujq,Ducloue:2019ezk}.
While resumming these logarithms is critical in order to gain full control over the calculations and achieve
 the desired NLO accuracy, this research direction  is largely irrelevant  to our  goals  
below and hence would not be reviewed in any further detail here. 

 The second group contains UV logs proportional to $\beta_0$, the one loop coefficient of the QCD $\beta$-function, which are naturally resummed into the running of the coupling constant $\alpha_s$. 
 Finally, there are potentially large
logs associated 
with the DGLAP evolution of either the projectile or the target (see e.g. \cite{Salam:1999cn}). More precisely, they are induced by the non-singular part of the DGLAP splitting functions at low x.

The main focus of this paper is  on the implementation of  running  coupling in the JIMWLK equation. 
The inclusion of  running coupling effects is known to be 
important from the phenomenological perspective: as mentioned, the 
running coupling BK  (rcBK) equation has been 
crucial 
in most phenomenological studies done so far \cite{Gotsman:2002yy,Albacete:2009fh,Albacete:2010sy,Albacete:2010bs,Kuokkanen:2011je,Albacete:2012xq,Lappi:2013zma,Iancu:2015joa,Albacete:2015xza,Ducloue:2019jmy,Beuf:2020dxl,Hanninen:2022gje}.
{
Since we are interested in the running coupling, our consideration of NLO JIMWLK Hamiltonian will be restricted to the $\beta_0$-dependent terms only (e.g. we omit the double logarithmic term in this paper).}

There exist several running coupling prescriptions,  which have been inferred from the calculation of 
NLO QCD diagrams. 
One prescription for rcBK is due to Balitsky \cite{Balitsky:2006wa} and another one, written both for rcBK and rcJIMWLK, is due to
Kovchegov and Weigert  (KW) \cite{Kovchegov:2006vj} (see also \cite{Gardi:2006rp}).  
The Balitsky prescription for rcBK can be uplifted to a prescription for rcJIMWLK as well. However, as we demonstrate below, 
these two rcJIMWLK Hamiltonians turn out to violate {\it positive semidefiniteness}. This means that their spectra 
have negative eigenvalues.   For some  particular observables %
this would lead to a wrong sign  of the energy evolution.  Instead, a fully consistent  high energy evolution Hamiltonian should have a non-negative spectrum.  Violation of positive semidefiniteness, 
as is observed both in the Balitsky and KW running coupling schemes, 
would lead to non-physical results for the solution of the corresponding evolution equation.

As will be explained later, the property of positive semidefiniteness of the Hamiltonian is tightly connected with the existence of a Langevin formulation of the evolution. A couple of running coupling prescriptions have been proposed in the literature for the Langevin formulation of JIMWLK, see e.g.~\cite{Lappi:2011ju,Lappi:2012vw}. They automatically correspond to positive semidefinite rcJIMWLK kernels. However, when expanded at NLO accuracy, none of these prescriptions   reproduce the terms commonly interpreted as running coupling corrections in the
NLO JIMWLK  Hamiltonian, following Refs.~\cite{Balitsky:2006wa} and \cite{Kovchegov:2006vj}. 

Our goal in this paper is to %
 review and asses various running coupling prescriptions. 
Ideally, a prescription for rcJIMWLK would lead to a positive semidefinite Hamiltonian, which would  also be compatible with fixed-order NLO results for the JIMWLK Hamiltonian.
For such a positive semidefinite rcJIMWLK Hamiltonian, we will explain how to obtain a Langevin 
formulation suitable for numerical simulations.
We find the running coupling prescriptions from Refs.~\cite{Balitsky:2006wa} and \cite{Kovchegov:2006vj}, as well as all the ones related to them that we could imagine, to lead  to a non positive semidefinite rcJIMWLK, which is thus unsuitable.

As has been realized in  \cite{KLS}, 
 at the level of the JIMWLK equation, 
the logarithms associated with the  DGLAP evolution of the projectile are proportional to $\beta_0$,
and can be confused with the ones associated with running coupling.
In Ref.~\cite{KLS}, a different choice of basis of color operators was used to write down the NLO JIMWLK Hamiltonian compared to Refs.~\cite{Balitsky:2006wa} and \cite{Kovchegov:2006vj}, 
in order to disentangle running coupling logs and DGLAP logs. This results in a different NLO contribution interpreted as running coupling correction than in the schemes of Refs.~\cite{Balitsky:2006wa} and \cite{Kovchegov:2006vj}.  In the scheme of Ref.~\cite{KLS}, the obtained running coupling correction at NLO leads by resummation to the daughter dipole prescription of Ref.~\cite{Lappi:2011ju}, corresponding to the replacement 
\beq
\alpha_s \mapsto \sqrt{\alpha_s(X^2)} \,\sqrt{\alpha_s(Y^2)}
\label{daughter}
\eeq
in the LO JIMWLK Hamiltonian, 
where $|X|$ ($|Y|$)  is the transverse size of the emitter-gluon pair in the wave-function (conjugate wave-function).
The daughter dipole prescription \eqref{daughter} is thus so far the only known running coupling prescription which is both compatible with the Langevin formulation of JIMWLK and consistent with NLO calculations {
albeit it requires further DGLAP resummation}. 

The paper is structured as follows. In Section \ref{Sec:LEforLO}, we review the stochastic reformulation of the LO JIMWLK evolution. In Section \ref{Sec:Formal_JIMWLK}, we discuss formal properties that should be obeyed by the JIMWLK evolution with a consistent running coupling prescription. In Section \ref{Sec:various_prescriptions}, we discuss various running coupling prescriptions available in the literature. Section \ref{Sec:discussions} presents a concise discussion of our analysis. Some technical details are presented in Appendix A and B.

\section{Langevin equation for fixed coupling LO  JIMWLK evolution}

\label{Sec:LEforLO}

In this section, %
we review the LO  JIMWLK evolution and its stochastic formulation. 
The Hamiltonian for the LO  JIMWLK evolution (see Ref.~\cite{Kovner:2005pe}, see \cite{Kovner:2013ona,Kovner:2014lca,Lublinsky:2016meo}) can be written in the following %
form 
\begin{equation}
\label{Eq:HLO}
H_{\rm LO}= \frac{\alpha_s}{2} \int_z  Q_i^a(z) Q_i^a(z)
\end{equation}
where $Q_i^a(z)$  has the interpretation of a single gluon emission operator\footnote{Throughout the manuscript, we use $\int_z \equiv \int d^2 z$ and $\int_k \equiv \int \frac{d^2 k}{(2\pi)^2}$ as shorthand notation for the coordinate and momentum space integrals over the transverse direction.}
\begin{equation}
Q_i^a(z)=  \frac{1}{\pi}\int_x K_i(x-z) \,q^a(x,z)
\,,\quad\quad\quad
q^a(x,z)=   [ U(x)  - U(z) ]^{ab}\,J_R ^b(x).
\label{def:q}
\end{equation}
The Weizsäcker–Williams (WW) field appearing in the single gluon emission operator is 
\begin{align}
    K_i(x) = \frac{x_i}{x^2}\,.
\end{align}
Additionally, $U(x)$ is the Wilson line along the light cone in the adjoint representation at the transverse position $x$. 
The fundamental Wilson line, $V(x)$, is related to the adjoint one through 
$$
U^{ab}(x) = 2 \, {\rm  tr}
\left [  V^\dagger(x) t^a  V(x) t^b \right]\,. 
$$
Here, $t_a$ are SU(N) generators in the fundamental  representation.  
The operators $J_R ^b(x)$ form a SU(N) algebra and act on the fundamental and adjoint  Wilson lines as right color rotations\footnote{Left rotations could be introduced similarly but are not required in this manuscript.}
 \begin{align}
    J_{R}^a (z) V(x) &= \delta^{(2)}(x-z) V(x)t^a,  \hspace{1cm}     J_{R}^a (z) V^\dagger(x) = -\delta^{(2)}(x-z) t^a V^\dagger(x)
\\
    J_{R}^a (z) U(x) &= \delta^{(2)}(x-z) U(x)T^a,  \hspace{1cm}    J_{R}^a (z) U^\dagger(x) =- \delta^{(2)}(x-z) T^a U^\dagger(x)\,,
\end{align}
where $T^a$ are SU(N) generators in the  adjoint representations.   

As a remark, note that the JIMWLK Hamiltonian \eqref{Eq:HLO} can be rewritten as
\begin{equation}
H_{\rm LO} = \int_{x,y,z}  {{\cal K}^{\rm LO}(x,y,z)} \,q^a(x,z)\, q^a(y, z)
\, ,\label{Eq:HLO_alt}
\end{equation}
with the kernel
\begin{equation}
{{\cal K}^{\rm LO}(x,y,z)} = \frac{\alpha_s}{2\pi^2}\, K_i(x-z)\, K_i(y-z)
=
\frac{\alpha_s}{2\pi^2}\, 
\frac{(x-z)_i}{(x-z)^2}\,
\frac{(y-z)_i}{(y-z)^2}
\,
\label{Eq:KLO}\, .
\end{equation}

Considering the rapidity variable $\eta$ as a fictitious time,  the rapidity evolution operator from $\eta_0$ to $\eta_1$ is 
\begin{equation}\label{evolop}
{\cal U}(\eta_0,\eta_1) = 
{\cal P}
e^{- \int_{\eta_0}^{\eta_1} d\eta H_{\rm LO}}\,.
\end{equation}
It was demonstrated in Refs.~\cite{Blaizot:2002np,Rummukainen:2003ns} that the evolution generated by (\ref{evolop}) is equivalent to a stochastic process 
of the Langevin type. Indeed, in its standard formulation the JIMWLK equation is a functional Fokker-Planck equation. As it is well known from statistical physics, there is a correspondence between Fokker-Planck and Langevin equations. JIMWLK equation is yet another example of this correspondence but functional equations.

The success of the Langevin reformulation lies in the observation
that $H_{\rm LO}$ is quadratic in the  operator $Q$. In the spirit of Hubbard–Stratonovich transformation,  to make  the evolution linear with respect to $Q$,
we introduce a local two-dimensional auxiliary vector field $\xi_i^a(\eta, x)$\footnote{Here and in the rest of the paper we will 
systematically ignore the Gaussian normalization constant, which is nevertheless always there.},    
\begin{align} 
\label{Eq:UHST}
\notag 
{\cal U}(\eta_0,\eta_1) & =  {\cal P}_{\eta}  \exp{- \int_{\eta_0}^{\eta_1} d\eta H_{\rm LO}} \\\notag  & =\,
   \int D \xi \,{\cal P}_{\eta}\, \exp{\int_{\eta_0}^{\eta_1} d\eta  \int_z  \left[ - i \sqrt{\alpha_s} Q_i^a(z)\xi^{a}_i(\eta, z)- \frac12 \vec{\xi}^{\,2}(\eta, z)\right]}\\ &=\,
  \int D\xi\; {\cal U}_\xi(\eta_0,\eta_1) \,e^{  - \int_{\eta_0}^{\eta_1} d\eta \int_z  \frac12 \vec{\xi}^{\,2}(\eta, z)} \,,\end{align}
where ${\cal P}_{\eta}$ corresponds to ordering of the $Q$ operators along $\eta$. 
Here, following the ideas borrowed
 from stochastic quantization,  we introduce an evolution operator ${\cal U}_\xi$  for a fixed configuration  of the field $\xi^{a}_i(\eta, x)$, 
 \begin{equation}
{\cal U}_\xi(\eta_0,\eta_1) = {\cal P}_{\eta}\,
 \exp{-i \int_{\eta_0}^{\eta_1} d\eta  \int_z    \sqrt{\alpha_s} Q_i^a(z)\xi_i^{a}(\eta,z)  }\,.
\end{equation} 
Then for an infinitesimally small rapidity interval $ \Delta $,
\begin{equation}
{\cal U}_\xi \equiv {\cal U}_\xi(\eta,\eta+\Delta) = 
 \exp{-i \Delta \int_z   \sqrt{\alpha_s} Q_i^a(z)\xi^{a}_i(\eta, z)  }\,.
\end{equation}
 From Eq.~\eqref{Eq:UHST}, we see that  
 the field $\xi$ is a random Gaussian field. 
For a stochastic process with  Gaussian noise, in order to account for all the effects linear in  $\Delta$, the 
evolution operator has to be expanded to the second order. 
This is simply due to the fact that 
\begin{align}
    \langle \xi^{a}_i  (\eta, x)  \xi^{b}_{j} (\eta', x') \rangle = \delta_{ij}\delta^{ab} \delta(\eta-\eta') \delta^{(2)}(x-x') 
\end{align}
which, after  discretization in  rapidity  $\eta_n=\eta_0 + n \Delta$, reads 
\begin{align}
    \langle \xi^{a\,,i}_{n} (x)  \xi^{b\,,j}_{n'} (x') \rangle = \frac{1}{\Delta} \delta^{ij}\delta^{ab} \delta_{n n'} \delta^{(2)}(x-x') \,.
\end{align}
Thus, it is convenient to rescale the noise variable as
$\hat \xi^{a\,i}_{n} (x) = \sqrt{\Delta}  \xi^{a\,i}_{n} (x) $
with variance which is independent of the step size $\Delta$.  In what follows, we drop 
the hat in the notation of the field $\hat \xi$. 
When expanded to the linear order in $\Delta$, the evolution operator reads 
\begin{align}
\label{Eq:Hexpand}
\notag
{\cal U}_\xi  &= 
 \exp{- i \sqrt{\alpha_s \Delta} \int_z    Q_i^a(z)\xi_{n}^{a,i}(z)  }
 \\&\approx 
 1 - i \sqrt{\alpha_s \Delta} \int_z    \xi_{n}^{a,i}(z)   Q_i^a(z)
 - \frac{1}{2} \alpha_s \Delta \int_z \int_{z'}  \xi_{n}^{a,i}(z)  \xi_{n}^{b,j}(z')
  Q_i^a(z)  Q_j^b(z')\,.
\end{align}
Note that if  the Gaussian integration over $\xi$ is performed, the linear term in $\xi$ drops while the quadratic term recovers the LO JIMWLK Hamiltonian as expected.  

Rapidity evolution of any operator that is built out of Wilson lines, such as a color dipole, 
is performed in two steps: first one computes evolution of the Wilson lines on a fixed configuration of the  noise and then averages the operator over the  noise. Thus, one can formulate the evolution of any operator by just tracking the evolution of a single  fundamental Wilson line. The action of the $Q$ operator on a fundamental Wilson line $V$ can be written  as 
\begin{align}
\label{Eq:QV}
\notag
    Q^a_i(z) V(x)  &= \frac{1}{\pi} K_i(x-z) [U(x)-U(z)]^{ab} V(x) t^b
     \\ & \notag =  \frac{1}{\pi} K_i(x-z) 
     V(x)  \left[ V^\dagger(x) t^a V(x)
     - V^\dagger(z) t^a V(z)\right]
     \\ &=  \frac{1}{\pi} K_i(x-z) 
     \left[ t^a 
     - V(x) V^\dagger(z) t^a V(z) V^\dagger(x) \right] V(x)\, ,
\end{align}
where the adjoint Wilson lines are written in terms of the fundamental ones.
Then, one step of the evolution of $V$ is obtained by substituting Eq.~\eqref{Eq:QV} into Eq.~\eqref{Eq:Hexpand}, 
\begin{align}
\label{Eq:OneStepAll}
{\cal U}_\xi V(x) &= 
V(x) - i  \frac{\sqrt{\alpha_s \Delta}}{\pi} \int_z 
K_i(x-z) \left[  \xi_{n}^{i}(z)  V(x)  - V(x) V^\dagger(z) \xi_{n}^{i}(z) V(z)   \right] \notag \\ & \quad \quad 
- \frac{\alpha_s \Delta}{2 \pi^2}
\int_z \int_{z'} K_i(x-z)K_j(x-z') \xi^{a\,, i}_{n}(z) \xi^{b\,, j}_{n}(z')
\notag \\ \notag & \quad \quad \times \bigg[ t^a t^b   V(x)  - t^a V(x) V(z')t^b V(z')
- t^b V(x) V(z)t^a V(z) \notag \\ & \hspace{6cm} + V(x)V^\dagger(z')t^b V(z') V^\dagger(z) t^a V(z)\bigg] \notag 
\\ & \quad \quad 
- \frac{\alpha_s \Delta}{2 \pi^2}
\int_z \int_{z'} K_i(x-z)K_j(z-z')  \xi^{a\,, i}_{n}(z) \xi^{b\,, j}_{n}(z') \notag \\ &\quad \quad  \times  \bigg[  V(x) V^\dagger(z) [t^b, t^a]_- V(z) 
 + V(x) [ V^\dagger(z)t^a V(z) , V^\dagger(z')t^b V(z')]_-
\bigg]  
\end{align}
where we denoted the commutators by $[A,B]_-$ to make the notation  more transparent. Additionally, we introduced $\xi^i_n(x)  = t^a \xi^{a,i}_n(x)$. 
 The commutator terms could be ignored when  averaged over the noise, since the latter produces a function symmetric in color,  
$\langle \xi^a \xi^b\rangle \sim \delta_{ab}$, while any correlation of $\xi$ with any Wilson line will lead to  contributions of order higher than linear in $\Delta$. 
With this nuance out of the way, we have  
\begin{align}
\label{Eq:OneStepall_simplified}
{\cal U}_\xi V(x) &= 
V(x) - i  \frac{\sqrt{\alpha_s \Delta}}{\pi} \int_z 
K_i(x-z) \left[  \xi_{i}(z)  V(x)  - V(x) V^\dagger(z) \xi_{i}(z) V(z)   \right] \notag \\ & 
- \frac{\alpha_s \Delta}{2 \pi^2}
\int_z \int_{z'} K_i(x-z)K_j(x-z') \xi^{a\,,i}_{n}(z) \xi^{b\,,j}_{n}(z') \notag
\\\ & \quad  \times
\bigg[ t^a t^b   V(x)  - t^a V(x) V(z')t^b V(z')
- t^b V(x) V(z)t^a V(z) \notag \\ & \hspace{6cm} + V(x)V^\dagger(z')t^b V(z') V^\dagger(z) t^a V(z)\bigg]
\end{align}
or simply 
\begin{align}
\label{Eq:OneStepall_simplified_2}
{\cal U}_\xi V(x) &=
V(x) - i  \frac{\sqrt{\alpha_s \Delta}}{\pi} \int_z 
K_i(x-z) \left[  \xi_{i}(z)  V(x)  - V(x) V^\dagger(z) \xi_{i}(z) V(z)   \right] \notag \\ & 
- \frac{\alpha_s \Delta}{2 \pi^2}
\int_z \int_{z'} K_i(x-z)K_j(x-z') \xi^a_{i}(z) \xi^b_{j}(z') 
\notag \\ & \quad \times
\bigg[ t^a t^b   V(x)  - 2 t^a V(x) V(z')t^b V(z')
 + V(x)V^\dagger(z')t^b V(z') V^\dagger(z) t^a V(z)\bigg]\,. 
\end{align}
It is straightforward to show that this expression can be  exponentiated into 
\begin{align}
\label{Eq:OneStepall_simplified_3}
{\cal U}_\xi V(x) &=
\exp \Bigg( - i   \frac{\sqrt{\alpha_s \Delta}}{\pi} \int_z 
\vec{K} (x-z) \cdot \vec{\xi}
_n(z) \Bigg) \notag \\ & \quad \times  V(x)
\exp \Bigg(  i   \frac{\sqrt{\alpha_s \Delta}}{\pi} \int_z 
\vec{K} (x-z) \cdot (V^\dagger(z) \vec{\xi}_n(z) V(z))\Bigg)\,. 
\end{align}
The main advantage of this re-exponentiated expression compared to \eqref{Eq:OneStepall_simplified_2}  is that it explicitly preserves  unitarity of the   Wilson line at each step of the evolution.
To reproduce the standard form of JIMWLK, one has to perform the rotation of the noise, see Ref.~\cite{Lappi:2012vw}, 
\begin{equation}
\label{Eq:rotate_noiseV}
    \tilde \xi(z) = V^\dagger(z) \xi(z) V(z) 
\end{equation}
which  leads to (renaming the noise again as $\xi$)
\begin{align}
\label{Eq:OneStepall_simplified_2_Lappi}
{\cal U}_\xi V(x) &=
\exp \Bigg( - i   \frac{\sqrt{\alpha_s \Delta}}{\pi} \int_z 
\vec{K} (x-z) \cdot (V(z) \vec{\xi}(z) V^\dagger(z))  \Bigg) V(x) \notag \\ & \quad \times 
\exp \Bigg(  i   \frac{\sqrt{\alpha_s \Delta}}{\pi} \int_z 
\vec{K} (x-z) \cdot  \vec{\xi}(z)\Bigg)\,. \end{align}
Note that the rotation does not change the variance of the noise.  To obtain \eqref{Eq:rotate_noiseV}, the
coupling of the noise to the operator $Q$ could be written in a slightly different but equivalent  way
\begin{equation}
{\cal U}(\eta_0,\eta_1) \,=\,
\int D\xi\; {\cal P}_{\eta} \,e^{\int_{\eta_0}^{\eta_1} d\eta  \int_z  [ - i \sqrt{\alpha_s} Q_i^a(z) { U^{ab}(z)}\xi^{b,i}(\eta, z)-\frac12 \vec{\xi}^{\,2}(\eta, z)]}\,.
\end{equation}

\section{Formal properties of JIMWLK with running coupling}

\label{Sec:Formal_JIMWLK}

Having reviewed the Langevin formalism for the fixed coupling in the previous section, we will focus on the running coupling case from now on. In this section, we will discuss general properties that should be obeyed by the JIMWLK evolution with a consistent prescription for running coupling.
In the next section, we will instead study specific running coupling prescriptions one by one.

\subsection{JIMWLK and BK beyond fixed coupling}

At NLO, new color structures appear in the JIMWLK Hamiltonian. But only NLO corrections involving the same color structure as the LO kernel can be interpreted as running coupling corrections to the LO JIMWLK kernel. Thus, any running coupling prescription for the improvement of the LO JIMWLK kernel can be written as
\begin{equation}
H = \int_{x,y,z}  {{\cal K}(x,y,z)} \,q^a(x,z)\, q^a(y, z)
\, ,\label{H_rc}
\end{equation}
with the same color operator $q^a(x,z)$ as defined in Eq.~\eqref{def:q}, and  
some yet unspecified %
kernel ${\cal K}(x,y,z)$ which encodes  the running coupling  effects. 
This kernel should reduce to the LO kernel \eqref{Eq:KLO} in the fixed coupling case. When expanded to the second order in $\alpha_s$, it should also reproduce the terms identified as running coupling terms within the NLO JIMWLK Hamiltonian, as will be discussed in the next section.

Given that a lot of discussion in the literature is focused on the BK equation, 
it is useful to remind the connection between the JIMWLK and BK evolutions, with either fixed or running coupling.  Consider the action of the JIMWLK Hamiltonian \eqref{H_rc} on a color dipole $S(u,v)=\tr[V(u)V^\dagger(v)]/N_c$. It is straightforward to obtain (see \cite{Kovner:2014lca})
the BK equation 
\begin{eqnarray}
\label{dipole}
H \,S(u,v)& =& N_c \int_z  K_{\rm dipole}(u,v,z)\,\left[ S(u,z) S(z,v) - S(u,v)   \right]\,, \\
\label{Kdipole}
K_{\rm dipole}(u,v,z) &\equiv &  {\cal K}(u,u,z) + {\cal K}(v,v,z) - {\cal K}(u,v,z) - {\cal K}(v,u,z)    \,.
\end{eqnarray}
The relation between the kernels can be  inverted to express ${\cal K}$ in terms  of $K_{\rm dipole} $ as
\begin{align}\label{dipoleinvert}
  {\cal K}(u,v,z) = - \frac12 K_{\rm dipole} (u,v,z) +f(u-z) +f(v-z)  
\end{align}
As this equation demonstrates, this inversion is not unique: it includes an arbitrary function $f$ of one transverse vector variable. 
This reflects  the fact that the JIMWLK evolution is more general and encodes more information than the BK evolution.
At LO, with fixed coupling, the relation between the JIMWLK and BK kernels is given by  Eqs.~(\ref{dipole}) and  (\ref{dipoleinvert}) with $f(X) = \frac{\alpha_s}{2\pi^2} \frac{1}{X^2}$.

\subsection{Positive semidefiniteness of the JIMWLK kernel}

The Hamiltonian $H$ must have all its eigenvalues 
non-negative. Otherwise, if the Hamiltonian has a negative eigenvalue, the relevant eigenstate will evolve with energy into a vacuum-like state that does not scatter. %

An alternative take on the JIMWLK evolution is  through its Fokker-Plank form  (see e.g. Ref.~\cite{Blaizot:2002np}), in which the energy evolution is viewed as a diffusion process. 
\begin{align}
\label{Eq:FP}
    \partial_\eta {\cal O}[\alpha] 
    = 
    &\frac12 \int d^2 x d^2 y 
    \frac{\partial} {\partial \alpha^{a}(x^-,x)} \left(   \eta^{ab}(x, y)      \frac{\partial} { \partial \alpha^{b}(y^-,y)}  {\cal O}[\alpha] 
    \right) 
\end{align}
with 
\begin{align}
    \eta^{ab} (x, y) = \int d^2 z {\cal K} (x,y,z) \left[ (U(x)-U(z))(U^\dagger(y)-U^\dagger(z)) \right]^{ab}
\end{align}
Here $\alpha$'s are phases of the Wilson lines.  
For Eq.~\eqref{Eq:FP} to have  interpretation of a diffusion process, the diffusion ``coefficient'' has to be positive semi-definite.  This must be true for arbitrary target  
(that is the configuration of the Wilson lines $U$) and arbitrary projectile 
(that can be thought as configuration of ${\delta\over\delta\alpha}$). 
This is equivalent to the positive semi-definiteness of the kernel  ${\cal K}(x,y,z)$. Hence below we will focus on the latter.

The property of positive semidefiniteness of the JIMWLK kernel is defined as follows. 
Regarded as an infinite matrix, the  kernel\footnote{Any kernel ${\cal K}(x,y,z)$ consistent with translational invariance can be written as a function of $X$ and $Y$ only.} ${\cal K}(x,y,z)={\cal K}(X\equiv x-z,Y\equiv y-z)$ is positive semidefinite if 
\begin{align}
\label{Eq:SPD}
    \int d^2X d^2 Y\, 
    \phi(X)\,{\cal K}(X,Y) \,\phi(Y) \,\ge\, 0\, 
\end{align}
for any function $\phi(X)$. 
This property is necessary  in order to avoid run-away/unstable solutions of the JIMWLK evolution Eq.~\eqref{Eq:JIMWLKonO}. For that reason, any kernel ${\cal K}(x,y,z)$ violating positive semidefiniteness \eqref{Eq:SPD} should be considered as unphysical. Moreover, that property is crucial in the derivation of the Langevin formulation, as will be discussed in the next subsection.

In order to explore consequences of the  positive semidefiniteness property  \eqref{Eq:SPD}, it is  convenient to restrict ourselves to trial functions of the form 
\begin{align}
    \label{Eq:phiX}
    \phi(X) = A_1 \delta(X-X_1) + A_2 \delta(X-X_2)\,.
\end{align} 
Then, the inequality \eqref{Eq:SPD} yields 
\begin{align}
\label{Eq:SPD2P}
    A_1^2 {\cal K}(X_1,X_1)  +  A_2^2 {\cal K}(X_2,X_2) 
    + 2  A_1 A_2  {\cal K}(X_1,X_2) 
    \ge 0 
\end{align}
for any $A_1$ and $A_2$. 
Note that the positivity of this quadratic form in $A_1$ and $A_2$ is a necessary but not sufficient condition for Eq.~\eqref{Eq:SPD}\footnote{A stronger condition can be obtained by extending the analysis to include $N>2$ points in Eq.~\eqref{Eq:phiX}. This will reduce to a positive-semidefiniteness of the corresponding quadratic form or an $N\times N$ matrix. A positive-semidefinite  matrix ${\cal K}$ with entries  ${\cal K}(X_i, X_j)$ can be written in the form 
$ {\cal K}  = R^T R$, where $R$ can be singular. }.
The relation \eqref{Eq:SPD2P} can be valid for any $A_1$ and $A_2$ only if
\begin{align}\label{cond1}
     &{\cal K}(X_1,X_1)  {\cal K}(X_2,X_2) 
    - {\cal K}^2(X_1,X_2) 
    \ge 0,   \\  
    &{\cal K}(X_1,X_1) \ge 0\ 
\label{cond1bis}
\end{align}
are satisfied
for any two coordinates $X_1$ and $X_2$. 

Moreover, by choosing $A_1 = - A_2 = A$ in the trial function, one arrives at the condition
\begin{align}
\label{Eq:cond3}
  {\cal K}(X_1,X_1)  +   {\cal K}(X_2,X_2) 
    - 2 {\cal K}(X_1,X_2) 
    \ge 0 \,.
\end{align}
Interestingly this inequality  is equivalent to
\begin{align}
  K_{\rm dipole} (X_1,X_2) 
    \ge 0 \,
    \label{Eq:cond3_dip}
\end{align}
as follows from Eq.~\eqref{Kdipole}. The inequality \eqref{Eq:cond3} is thus a necessary (but not sufficient) condition for the positive semidefiniteness of the JIMWLK Hamiltonian. It is a less restrictive condition than Eq.~\eqref{cond1}.

In the next section, we will check the validity of the inequalities \eqref{cond1}, \eqref{cond1bis} and \eqref{Eq:cond3} for various prescriptions that can be proposed for rcJIMWLK, as a test for the positive semidefiniteness of the corresponding kernels.

As an example, the inequality \eqref{cond1} can be verified for the LO JIMWLK kernel \eqref{Eq:KLO} as follows:
\begin{align}
\notag  
     {\cal K}^{\rm LO}(X_1,X_1)  {\cal K}^{\rm LO}(X_2,X_2) 
    - {{\cal K}^{\rm LO}}(X_1,X_2)^2  &=
    \left( \frac{\alpha_s}{2\pi^2} \right)^2
    \left [ 
    \frac{1}{X_1^2 X_2^2} - 
    \frac{(X_1\cdot X_2)^2 }{X_1^4 X_2^4} 
    \right]  \\ & 
    =  
    \left( \frac{\alpha_s}{2\pi^2} \right)^2  \frac{(X_1\times X_2)^2 }               {X_1^4 X_2^4}  \ge 0 
\label{pos_semidef_LO}
\end{align}
where $ X_1\times X_2  = \epsilon_{ij} (X_1)_i  (X_2)_j  $. 

\subsection{Langevin formulation for a positive semidefinite JIMWLK evolution 
}
Assuming that a positive semidefinite kernel is available, one can construct the corresponding Langevin formulation in the following way. 
For any positive semidefinite kernel, it is possible to define a ``square root'' $\Psi$, such that
\begin{align}\label{Psi}
\frac{1}{2}\int_V    \Psi_i(X,V)
    \Psi_i(Y,V) = {\cal K} (X,Y)\,.
\end{align}
Here, the subscript $i$ demonstrates that, in general, the field $\Psi$ can be a vector. The equations for a scalar field $\Psi$ can be obtained by ignoring the subscript.  
In general, finding  $\Psi_i(X,V)$ 
is a rather nontrivial process.   However, this can be achieved by performing a Schur decomposition since the Langevin equation is formulated on a discretized space.
Once $\Psi_i(X,V)$ is known, the corresponding Langevin equation can be obtained by following the same steps discussed in Sec.~\ref{Sec:LEforLO}. Namely, 
one introduces a
bi-local scalar noise field $\zeta$, such that for a small rapidity interval $\Delta$, the evolution operator reads\\
\begin{align}
\label{Eq:UdelBi}
{\cal U} (\eta, \eta + \Delta ) 
\,=\,
&\int D\zeta
      e^{ {\Delta}\, \left(
    i \int_{x,y,z} \Psi_i(X,Y) \,q^a(x,z) \,\zeta_i^a(y,z) - \frac{1}{2} \int_{x,z}  \zeta_i^a(x,z)\zeta_i^a(x,z)
    \right)
    }\,.
\end{align}
Indeed, integrating out the noise field $\zeta_i^a(x,z)$ yields 
\begin{align}
{\cal U}(\eta, \eta + \Delta )   \,=\,
&   e^{ - {\Delta}\, 
     \int_{x,y,z} {\cal K}(X,Y) \,q^a(x,z) \,q^a(y,z) 
    }\,.
\end{align}
Following the same steps introduced in Sec.~\ref{Sec:LEforLO} we obtain the following Langevin equation:
\begin{align}
\label{Eq:BiLocal}
{\cal U}_\zeta V(x) &=
\exp \Bigg( - i  \sqrt{\Delta}  \int_{y,z} 
\vec{\Psi} (X, Y) \cdot (V(z) \vec{\zeta}(y,z) V^\dagger(z))  \Bigg) V(x) \notag \\ & \quad \times 
\exp \Bigg(  i  
\sqrt{\Delta}  \int_{y,z} 
\vec{\Psi} (X, Y)  \vec{\zeta}(y,z)  
\Bigg)\,. \end{align}
For ${\cal K}={\cal K}^{\rm LO}$, it is straightforward to demonstrate that the bi-local noise can be equivalently replaced with a local noise recovering the results of Section 2, see Appendix~\ref{App:bitosi}.

\section{Analysis of the available prescriptions for rcJIMWLK}

\label{Sec:various_prescriptions}

\subsection{Scheme dependence of running coupling prescriptions\label{sec:scheme_dep}}

In theories like QCD and QED, adopting a running coupling prescription amounts to performing the all order resummation of a set of potentially large logarithms (which are leftovers after the UV renormalization procedure) in a perturbative series, thus improving the accuracy compared to strictly fixed order results. This resummation is very well understood and under control in QED. By contrast, in QCD, there is a significant scheme-dependence associated with such resummation, due notably to our poorer understanding of perturbative series at high orders. %
There are three main sources of scheme-dependence for running coupling prescriptions in QCD, associated with three successive steps in the construction of a running coupling prescription:
\begin{enumerate}
\item proper definition of the considered observable, 

\item identification of running coupling correction among higher order corrections, 

\item resummation of the identified running coupling corrections.
\end{enumerate}

Most of the literature about running coupling prescriptions in QCD in general focuses on the second and third points, because well defined observables (like cross sections for simple processes) are considered. By contrast, in the case of the JIMWLK evolution, the first point is crucial, for the following reason.
As is well known, new color operators arise at each order in the perturbative expansion of the JIMWLK Hamiltonian. In particular, some higher order diagrams are UV divergent even though they involve new color operators. At NLO, this is the case for diagrams in which a gluon splits into a gluon-gluon or a quark-antiquark pair, which then interacts with the target shockwave, and recombine later into a gluon. These diagrams are UV divergent in the limit of small gluon or quark-antiquark loop. However, due to the locality of the QCD counterterms, the UV divergence for such diagrams cannot be removed by a contribution containing the same color operator. Instead, one should use the property of color coherence of QCD: in the limit of small gluon or quark-antiquark loop, the loop is not resolved by the target, and thus has the same interaction with the target as its parent gluon. For that reason, even if these NLO diagrams are UV divergent and involve new color operators not present in the LO JIMWLK Hamiltonian, their UV divergences  only involve the color operator present at LO. The strategy is then to carefully choose a basis of color operators to express the NLO JIMWLK Hamiltonian, which contains, in addition to the LO color operator, the new operators with appropriate subtractions by the LO color operator.
In such a way, all the UV divergences are transferred to the coefficient of the color operator already present at LO, and UV renormalization can be performed with the standard  QCD counterterms (for example in the $\overline{\rm MS}$ scheme).

The choice of basis of color operators in this procedure is however not unique, due to the possibility to subtract UV finite contributions together with the UV divergences in the definition of the new color operators, which will lead to a different coefficient at NLO for the LO color operator. For example, in Ref.~\cite{Balitsky:2006wa}, Balitsky  performed the UV subtraction of the new color operator associated to a quark-antiquark loop with a term in which the parent gluon Wilson line is placed at the position of the quark. By contrast, in Ref.~\cite{Kovchegov:2006vj}, Kovchegov and Weigert performed the UV subtraction of the new color operator associated to a quark-antiquark loop with a term in which the parent gluon Wilson line is instead placed at the center of the quark-antiquark pair (weighted by light-cone momentum fractions). 
These two choices of UV subtraction scheme, or equivalently of basis of color operators, lead to a different NLO correction to the coefficient of the LO color operator, and running coupling prescriptions for JIMWLK based on either choice will typically differ for that reason. See Ref.~\cite{Albacete:2007yr} for a quantitaive study of that scheme dependence. A priori, there is no compelling argument for preferring one UV subtraction scheme over the other. Moreover, infinitely many other UV subtraction schemes could be constructed and used, instead of the Balitsky~\cite{Balitsky:2006wa} or the Kovchegov-Weigert~\cite{Kovchegov:2006vj} UV subtraction schemes. 

In order to formulate a running coupling improvement of the LO JIMWLK kernel \eqref{Eq:KLO}, the first scheme dependence that one  encounters is thus associated with the proper definition of that kernel beyond LO, which requires to specify an appropriate basis of color operators to write the JIMWLK Hamiltonian beyond LO, corresponding to the choice of an UV subtraction scheme.  

Once such a choice of basis is made, and corresponding  NLO corrections are calculated,
 the second step is to decide which part of the NLO corrections to this kernel should be associated with running coupling effects. This choice is usually performed by selecting terms proportional to the coefficient $\beta_0$ of the QCD $\beta$  function, obtained via their dependence on the number of active quark flavors $N_f$ \cite{Brodsky:1982gc}. Note however that this step can sometimes be ambiguous as well, due to the possible extra dependence of observables on $N_f$ at higher orders, beyond the $N_f$ dependence via $\beta_0$. This strategy can be generalized to higher orders. For example, one could identify running coupling corrections among NNLO corrections by selecting terms with coefficients ${\beta_0}^2$, $\beta_1$ and $\beta_0$. 

Finally, once running coupling terms are identified among the higher order corrections, the third and last step is to resum them, in order to obtain a running coupling prescription, as stated above. 
In QED, this step corresponds to the summation of a geometric series. In QCD, this is not exactly the case, and perturbative series are not well understood at high orders. For that reason, various models have been proposed in the literature in order to write a resummed expression from the knowledge of the first few fixed-order corrections, but so far there is no consensus on a best resummation procedure. This leads to a resummation scheme ambiguity for running coupling prescriptions in QCD.
This resummation scheme ambiguity can be in principle reduced by including the results from higher and higher fixed order calculations. But an ambiguity would always remains, due to the practical impossibility to obtain genuine all order results in perturbative QCD.

All in all, in the construction of a rcJIMWLK kernel from the NLO Hamiltonian, there are three types of scheme dependence, associated with the three steps in the general procedure:
\begin{enumerate}
\item UV subtraction scheme ambiguity,

\item ambiguity in the identification of the $\beta_0$ terms, 

\item resummation scheme ambiguity.
\end{enumerate}

\subsection{NLO fixed-order JIMWLK kernel in the Balitsky scheme}

At NLO, the JIMWLK Hamiltonian can be written as~\cite{Kovner:2013ona,Lublinsky:2016meo} 
\begin{equation}
H_{\rm NLO} = \int_{x,y,z}  {{\cal K}_{JSJ}(x,y,z)}\; q^a(x,z)\, q^a(y, z)\, +\ldots
\end{equation}
with the explicitly written term having the same color structure as the LO contribution, and extra NLO corrections involving new color operators not explicitly written.
Hence,
\begin{align}
    {\cal K}_{JSJ}(x,y,z)  = {\cal K}^{\rm LO} (x,y,z)+ {\cal K}^{\rm NLO}_{JSJ}(x,y,z)\, ,
    \label{KJSJ_LO_plus_NLO}
\end{align}
with the leading order kernel \eqref{Eq:KLO}
that explicitly factorizes in $x$ and $y$ resulting in the representation \eqref{Eq:HLO}. 

As discussed in the previous subsection, the kernel ${\cal K}_{JSJ}^{\rm NLO}(x,y,z)$ is scheme dependent: its value depends on the chosen basis of color operators to write the JIMWLK Hamiltonian. 
Following Refs.~\cite{Kovner:2013ona,Lublinsky:2016meo} in which the NLO JIMWLK Hamiltonian was derived, one can adopt the Balitsky scheme~\cite{Balitsky:2006wa}
to perform the 
UV subtraction of the new color operators, and thus to specify the chosen basis of color operators.
Using that scheme to define ${\cal K}_{JSJ}^{\rm NLO}$, one finds
\footnote{
In Eq.~(\ref{Eq:JIMWLKBal_fixed_order_NLO}) and throughout the paper we omitted potentially large, but  $\beta_0$-independent  logarithms (including the double logarithmic terms), as they are  irrelevant for the discussion of running coupling and are outside the scope of this paper.}
\begin{align}
{\cal K}_{JSJ}^{\rm NLO}(x,y,z)=&\,
{\alpha_s^2\over 16\pi^3}\, \beta_0\, \left\{
-  \frac{(x-y)^2}{X^2 Y^2}\ln(x-y)^2\hat{\mu}^2
\right. 
\nonumber\\  
& \hspace{1.65cm} \left. 
+ \frac{X^2-Y^2}{ X^2 Y^2}\ln\frac{X^2}{Y^2} + 
\frac{1}{X^{2}}\ln X^{2}\hat{\mu}^{2}+\frac{1}{Y^{2}}\ln Y^{2}\hat{\mu}^{2} 
\right\} 
\nonumber\\
& 
+{\alpha_s^2\over 8\pi^3}\,
\frac{X\!\cdot\! Y}{X^{2}Y^{2}}\,
\left\{ \left(\frac{67}{9}\!-\!\frac{\pi^2}{3}\right)N_c - \frac{10}{9}\, N_f\right\}
\label{Eq:JIMWLKBal_fixed_order_NLO}
\end{align}
after performing UV renormalization, using conventional dimensional regularization and the $\overline{\rm MS}$ scheme. Hence, $\alpha_s$ is now the renormalized coupling in this scheme, implicitly taken at the $\overline{\rm MS}$ scale $\mu_{\overline{\rm MS}}^2$. We have introduced the notation
\begin{equation}
\hat{\mu} \equiv \frac{1}{2}\, e^{\gamma_E}\, \mu_{\overline{\rm MS}}
\, ,
\label{def_mu_hat}
\end{equation}
where $\gamma_E$ is the Euler–Mascheroni constant. The first coefficient $\beta_0$ of the QCD beta-function is normalized as
\begin{equation}
\beta_0 = \frac{11}{3}\, N_c- \frac{2}{3}\, N_f
\, .
\label{def_beta_0}
\end{equation}

The kernel ${\cal K}_{JSJ}^{\rm NLO}(x,y,z)$ from Eq.~\eqref{Eq:JIMWLKBal_fixed_order_NLO} contains two terms. In the first term, corresponding to the first two lines, the contributions from gluon loop diagrams (proportional to $N_c$) and the contributions from quark loop diagrams (proportional to $N_f$) have combined in such a way that an overall factor $\beta_0$ is obtained. This term has a non-trivial dependence on the coordinates, involving in particular single logarithms. That whole term is then typically interpreted as  being  associated  with running coupling.

By contrast, in the second term in Eq.~\eqref{Eq:JIMWLKBal_fixed_order_NLO}, corresponding to  the third line, the quark and gluon diagrams contributions are not fully combining into a $\beta_0$ factor. One could still single out a piece proportional $\beta_0$ out of it, and interpret it as a contribution associated with running coupling, but there is no unique way to do it. This is precisely the second type of scheme dependence for running coupling prescriptions described in the previous subsection. However, that second term in Eq.~\eqref{Eq:JIMWLKBal_fixed_order_NLO} is proportional to ${\cal K}^{\rm LO}(x,y,z)$, up to a constant factor. Hence, it cannot change the qualitative properties of the JIMWLK kernel, by contrast to the first term in Eq.~\eqref{Eq:JIMWLKBal_fixed_order_NLO}. One could also entirely eliminate this second term by switching the UV renormalization scheme from the $\overline{\rm MS}$ scheme to the gluon bremsstrahlung scheme~\cite{Catani:1990rr,Dokshitzer:1995ev}. Or one could simply leave that second term as a harmless contribution to the NLO JIMWLK Hamiltonian, not to be resummed into a running coupling prescription. For these reasons, we will focus from now on the first term Eq.~\eqref{Eq:JIMWLKBal_fixed_order_NLO} only, proportional to $\beta_0$, and discard the second term.

Note that at NLO accuracy, the kernel ${\cal K}_{JSJ}(x,y,z)$ does not factorize anymore, preventing us to write a simple generalization of the expression \eqref{Eq:HLO} for the JIMWLK Hamiltonian beyond LO.

Before we proceed with the resummation into a running coupling prescription, it is instructive to check if the fixed order NLO kernel ${\cal K}_{JSJ}(x,y,z) $ in Balitsky's UV subtraction scheme, written in Eqs.~\eqref{KJSJ_LO_plus_NLO} and \eqref{Eq:JIMWLKBal_fixed_order_NLO}, is positive semi-definite, or at least if the inequality \eqref{cond1} is valid. 
Keeping terms up to  $\alpha^3_s$ terms only (since higher orders are not under  control), one finds  
\begin{align}
\label{Eq:KpsitivityExp}
    {\cal K}_{JSJ}(X,X)& {\cal K}_{JSJ}(Y,Y) - {\cal K}^2_{JSJ}(X,Y) 
    = 
    {\cal K}^{\rm LO}(X,X) {\cal K}^{\rm LO}(Y,Y) - \left[{\cal K}^{\rm LO}(X,Y) \right]^2 \notag 
    \\ & +   {\cal K}^{\rm LO}(X,X) {\cal K}^{\rm NLO}_{JSJ}(Y,Y)
    +   {\cal K}^{\rm NLO}_{JSJ}(X,X) {\cal K}^{\rm LO}(Y,Y)
    \notag \\ & - {\cal K}^{\rm LO}(X,Y) {\cal K}^{\rm NLO}_{JSJ}(X,Y)
    -  {\cal K}^{\rm NLO}_{JSJ}(X,Y) {\cal K}^{\rm LO}(X,Y)
    +O(\alpha^4_s)
\end{align}
First, to simplify our task, we consider the particular case of collinear vectors, $Y=c X$.  
This choice completely eliminates  the contribution of order $\alpha^2_s$, that is 
\begin{align}
\label{Eq:ColinearLO}
{\cal K}^{\rm LO}(X,X) {\cal K}^{\rm LO}(cX,cX) - \left[{\cal K}^{\rm LO}(X,cX) \right]^2 = 0\,.     
\end{align}
The LO kernels are 
\begin{align}
    {\cal K}^{\rm LO}(X,X) = \frac{\alpha_s}{2\pi^2} \frac{1}{X^2}  \,,~~~~~
    {\cal K}^{\rm LO}(c X,c X) = \frac{\alpha_s}{2\pi^2} \frac{1}{c^2 X^2}  \,, ~~~~~
    {\cal K}^{\rm LO}( X,c X) = \frac{\alpha_s}{2\pi^2} \frac{1}{c X^2}  \,
\end{align}
while the NLO kernels read (keeping only the $\beta_0$ contribution, since the other contribution, proportional to ${\cal K}^{\rm LO}$, will drop at this order due to Eq.~\eqref{Eq:ColinearLO}) 
\begin{eqnarray}
{\cal K}_{JSJ}^{\rm NLO}(X,X ) &= &
{\alpha_s^2\over 16\pi^3}
\frac{2 \beta_0}{X^{2}}\ln X^{2}\hat{\mu}^{2}, ~~~~~~~~
{\cal K}_{JSJ}^{\rm NLO}(c X,c X ) =
{\alpha_s^2\over 16\pi^3}
\frac{2 \beta_0}{c^2 X^{2}}\ln c^2 X^{2}\hat{\mu}^{2}\nonumber \\\nonumber  && \\
{\cal K}_{JSJ}^{\rm NLO}(X, c X)&=&
{\alpha_s^2\over 16\pi^3} \left\{
- \beta_0 \frac{(1-c)^2}{c^2 X^2}\ln(1-c)^2 X^2\hat{\mu}^2
\right. \\   
&+& \left.\beta_0 \frac{1-c^2}{ c^2 X^2}\ln\frac{1}{c^2} + 
\frac{\beta_0}{X^{2}}\ln X^{2}\hat{\mu}^{2}+\frac{\beta_0}{c^2 X^{2}}\ln c^2 X^{2}\hat{\mu}^{2} 
\right\} \,.\nonumber
\end{eqnarray}
Substituting these into 
Eq.~\eqref{Eq:KpsitivityExp} yields 
\begin{multline}
     \label{Eq:KpsitivityExp1}
    {\cal K}_{JSJ}(X,X) {\cal K}_{JSJ}(c X,c X) - {\cal K}^2_{JSJ}(X,c X) 
    \\ =  \frac{\alpha_s^3\, \beta_0}{16 \pi^5} 
    \frac{(1\!-\!c)}{c^3 X^4}
\Big[   (1\!-\!c) \ln ((1\!-\!c)^2)+ c \ln( 
 c^2) \Big]  +O(\alpha^4_s)\,. %
\end{multline}
The function on the right-hand side of 
(\ref{Eq:KpsitivityExp1}) vanishes at $c=1$. It is now a simple exercise to check that this function is negative for values of $c$ different from $0$ and $1$ (and goes to $-\infty$ at $c=0$).

This proves that the fixed order NLO kernel ${\cal K}_{JSJ}(x,y,z)$ in Balitsky scheme for UV subtraction violates
 the inequality \eqref{cond1} at least for some values of the coordinates $X$ and $Y$. Hence, it violates %
 the condition of positive semidefiniteness \eqref{Eq:SPD}.

 For completeness, let us also discuss the conditions \eqref{cond1bis} and \eqref{Eq:cond3} for the JIMWLK and BK kernels at fixed NLO order with Balitsky's scheme UV subtraction.
From the LO kernel \eqref{Eq:KLO}, and the $\beta_0$ part of Eq.~\eqref{Eq:JIMWLKBal_fixed_order_NLO}, one finds
\begin{align}
{\cal K}_{JSJ}(X,X ) = &\,
\frac{\alpha_s}{2\pi^2} \frac{1}{X^2}
\left[1+
{\beta_0\,\alpha_s\over 4\pi}
\,\ln \left(X^{2}\hat{\mu}^{2}\right)
\right]
\, .\label{test_cond1bis_NLO_Bal}
 \end{align}
On the one hand, in the limit $\alpha_s\rightarrow 0$, the expression \eqref{test_cond1bis_NLO_Bal} is dominated by the LO term, which is always positive, so that the condition \eqref{cond1bis} is satisfied in that limit.
On the other hand, for any finite value of $\alpha_s$ (and the corresponding value of $\mu_{\overline{\rm MS}}$), one finds from the expression \eqref{test_cond1bis_NLO_Bal} the equivalence
\begin{align}
{\cal K}_{JSJ}(X,X ) \geq 0 &\, \quad
\Leftrightarrow \quad
X^2 \geq \frac{1}{\hat{\mu}^2}\,
e^{-\frac{4\pi}{\beta_0\, \alpha_s}}
\, .\label{test_cond1bis_NLO_Bal_2}
 \end{align}
Note that in practice, that lower bound  for $X^2$ corresponds to an \emph{extremely} small distance, due to the essential singularity for $\alpha_s\rightarrow 0$. Hence, the inequality \eqref{cond1bis} is valid in most of the available range for $X$, but violated at extremely short distances. This is a further proof that the JIMWLK kernel at fixed NLO order with Balitsky's scheme UV subtraction is not positive semidefinite. The condition of positivity of the dipole kernel \eqref{Eq:cond3} can be studied in a similar way, and similar conclusions can be found. On the one hand, the dipole kernel is positive in the $\alpha_s\rightarrow 0$ limit, due to the dominance of the LO contribution. But for any finite value of $\alpha_s$ (or $\mu_{\overline{\rm MS}}$), that kernel becomes negative in specific regions in $X$ and $Y$, in which large negative logs arise in the NLO contribution and overcome the LO contribution.

The large logs multiplied by $\beta_0$ leading to the violation of the conditions \eqref{cond1bis} and \eqref{Eq:cond3} are precisely the ones that running coupling presciptions are supposed to resum.
Hence, the violation of the conditions \eqref{cond1bis} and \eqref{Eq:cond3} for the JIMWLK and BK kernels at fixed NLO order simply emphasizes the need to use running coupling instead of pure fixed order in practice. 
Then, one can expect that the conditions \eqref{cond1bis} and \eqref{Eq:cond3} are  restored by resumming running coupling effects. 
Hence, let us now study the JIMWLK kernel with Balitsky's prescription for running coupling.

\subsection{Balitsky's prescription for rcJIMWLK}

The dipole kernel in Balitsky's running coupling prescription~\cite{Balitsky:2006wa} reads 
\begin{align}
K_{\rm dipole}^{\rm B} (x,y,z) 
&= 
\frac{\alpha_s((X-Y)^2)}{2\pi^2} \left[ 
-2 \frac{X\cdot Y} {X^2 Y^2} + \frac{\alpha(X^2)}{\alpha_s(Y^2)} \frac{1}{X^2}
+ \frac{\alpha_s(Y^2)}{\alpha_s(X^2)} \frac{1}{Y^2}
\right]\,
\label{Eq:dipole_kernel_Bal}
\end{align}
where we adopted an abbreviated notation for the coupling $\alpha_s$ at a squared momentum scale $\mu_X^2$ constructed from a spacial distance $|X|$
\begin{equation}
\alpha_s(X^2) \equiv \alpha_s\left (\mu_X^2 = \frac{4}{X^2} e^{-2 \gamma_E}  \right)
\, .
\label{def_alpha_s_pos_space}
\end{equation}
Using the relation between the dipole kernel and the JIMWLK kernel given in Eq.~\eqref{dipoleinvert}, we obtain 
\begin{multline}
{\cal K}^{\rm B} (x,y,z) 
= 
\frac{\alpha_s((X-Y)^2)}{2\pi^2} \left[ 
\frac{X\cdot Y} {X^2 Y^2} -  \frac{\alpha(X^2)}{2 \alpha_s(Y^2)} \frac{1}{X^2}
-  \frac{\alpha_s(Y^2)}{2 \alpha_s(X^2)} \frac{1}{Y^2}
\right] +  f(X) + f(Y)  \,.     
\end{multline}
As discussed  previously, the function $f$ is undetermined in the inverse relation~\eqref{dipoleinvert}. Here however, it can be fully fixed by the requirement that the perturbative expansion of ${\cal K}^{\rm B} (x,y,z)$ has to reproduce Eq.~\eqref{KJSJ_LO_plus_NLO} with the $\beta_0$ contribution from Eq.~\eqref{Eq:JIMWLKBal_fixed_order_NLO}
at NLO accuracy. This yields
\begin{align}
    \label{Eq:JIMWLKBal}
    {\cal K}^{B}(x,y,z) = \frac{\alpha_s((X-Y)^2)}{2\pi^2} \frac{X \cdot Y}{X^2 Y^2} 
    & + 
    \frac{\alpha_s(X^2)}{4 \pi^2} \frac{1}{ X^2} \left(1 - \frac{\alpha_s((X-Y)^2)}{\alpha(Y)} \right)
    \notag  \\ & + 
    \frac{\alpha_s(Y^2)}{4 \pi^2} \frac{1}{ Y^2} \left(1 - \frac{\alpha_s((X-Y)^2)}{\alpha(X^2)} \right)\,. 
\end{align}
Indeed, the expression \eqref{Eq:JIMWLKBal} can be expanded at NLO accuracy, thanks to the relation 
\begin{align}    
\alpha_{s}(X^2) 
= &\, 
\alpha_s + \frac{\alpha_s^2 \beta_0}{4\pi} \ln \left(X^2 \hat{\mu}^2\right)
+O(\alpha_{s}^3)
\label{alpha_s_expand}
\, ,
\end{align}
in which the coupling is always taken at the scale $\mu_{\overline{\rm MS}}^2$ on the right hand side, and the notations \eqref{def_mu_hat} and
\eqref{def_alpha_s_pos_space} were used.

We now check if the resulting rcJIMWLK kernel is positive semidefinite. 
Due to asymptotic freedom, $\alpha_s((X\!-\!Y)^2) \to 0$ for $(X\!-\!Y)^2 \to 0$. Using that property, one finds from Eq.~\eqref{Eq:JIMWLKBal} that
\begin{align}
    {\cal K}^{B}(x,x,z) \equiv  {\cal K}^{B}(X,X)
=    \frac{\alpha_s(X^2)}{2 \pi^2} \frac{1}{ X^2} \,.
\label{Eq:JIMWLKBal_XX} 
\end{align}
The condition \eqref{cond1bis} is thus valid if the coupling $\alpha_s(X^2)$ is always positive. On the one hand, the coupling should indeed be positive for the consistency and stability of the theory. On the other hand, the one-loop expression for the coupling\footnote{At one loop, the scale $\Lambda$ corresponding to the Landau pole is obtained as
$\Lambda^2=\mu_{\overline{\rm MS}}^2\, e^{-\frac{4\pi}{\beta_0\, \alpha_s}} $. It is independent of $\mu_{\overline{\rm MS}}$ at that accuracy, since the implicit dependence on $\mu_{\overline{\rm MS}}$ through $\alpha_s$ compensates the explicit dependence. }, 
\begin{align}
   \alpha_s(X^2) = \frac{4\pi}{\beta_0 
   \ln \left(\frac{\mu_X^2}{\Lambda^2}\right)}
    = \frac{4\pi}{\beta_0 
   \ln \left(\frac{4\, e^{-2 \gamma_E}}{X^2\Lambda^2}\right)}
   \, ,
   \label{1loop_alpha_s}
\end{align}
is positive in the perturbative UV domain but diverges at $\mu_X^2=\Lambda^2$ and then goes negative further in the IR domain. In phenomenological studies, the one-loop expression \eqref{1loop_alpha_s} is often modified in the IR in order to keep $\alpha_s(X^2)$ bounded and positive, for example by imposing a finite IR limit for $\alpha_s $, also known as freezing coupling prescription. In such a case, $\alpha_s(X^2)$ is always positive, so that ${\cal K}^{B}(X,X)>0$ for  all $X$, which corresponds to the condition \eqref{cond1bis}.
Moreover, we note that the dipole kernel \eqref{Eq:dipole_kernel_Bal} can be rewritten in the form
\begin{align}
\notag
K^B_{\rm dipole} (x,y,z) 
& =  
\frac{\alpha_s((X-Y)^2)}{2\pi^2 \alpha_s(X^2) \alpha_s(Y^2)} \frac{1}{X^2 Y^2} 
\left[ \alpha_s(X^2) Y  - \alpha_s(Y^2) X  \right]^2  \,.
\end{align}
Hence, provided that the coupling is always positive, thanks to the use of such prescription in the IR, one has
$K^B_{\rm dipole} (x,y,z)\ge 0$ for all values of $x,y,z$, which correspond to the condition \eqref{Eq:cond3}. Therefore, both conditions \eqref{cond1bis}
and \eqref{Eq:cond3}, which were violated in the case in the case of the fixed order NLO kernel (with Balitsky's UV subtraction scheme), are restored by the running coupling resummation, provided the coupling is prevented to go negative in the IR. 

Let us now discuss the fate of the nonlinear condition \eqref{cond1} in the case of Balitsly prescription \eqref{Eq:JIMWLKBal} for the rcJIMWLK kernel. 
Let us focus on the case of $Y=c X$ for simplicity, and more precisely on the regime of $c\rightarrow 1$, meaning that $|x\!-\!y|\ll|x\!-\!z|\sim |z\!-\!y|$. We assume moreover that $X$ is small enough, in the perturbative regime, so that the one-loop expression \eqref{1loop_alpha_s} for the running coupling is valid.
Then, one can perform the Taylor expansion of the coupling $\alpha_s(Y^2)$ as
\begin{align}
\alpha_s(Y^2) =  \alpha_s(c^2 X^2)
= 
\alpha_s(X^2)\Bigg\{
1
&\,
-\frac{\beta_0 \alpha_s(X^2)}{2\pi}
(1\!-\!c)
+O((1\!-\!c)^2)
\Bigg\}
\label{aY_expand_c1}
\end{align}
for $c\rightarrow 1$. By contrast, $\alpha_s((X-Y)^2)$ goes to $0$ logarithmically in that limit, as
\begin{align}
\alpha_s((X-Y)^2) =  \alpha_s((1\!-\!c)^2 X^2)
\sim &\,
\frac{4\pi}{\beta_0}\,
\frac{1}{\ln\left(\frac{1}{(1\!-\!c)^2}\right)}
\rightarrow 0
\, .
\end{align}
Using Eq.~\eqref{aY_expand_c1}, one finds the Taylor expansion
\begin{align}
{\cal K}^{B}(Y,Y)
=&\, {\cal K}^{B}(cX,cX)
=\frac{\alpha_s(X^2)}{2 \pi^2} \frac{1}{ X^2}
\Bigg\{
1
+2(1\!-\!c)\left[1\!-\!\frac{\beta_0 \alpha_s(X^2)}{4\pi}\right]
+O((1\!-\!c)^2)
\Bigg\}
\label{K_Bal_YY_expand_c1}
\, .
\end{align}
By contrast, the dipole kernel \eqref{Eq:dipole_kernel_Bal} becomes
\begin{align}
K_{\rm dipole}^{\rm B} (X,cX)
=&\,
\frac{\alpha_s((1\!-\!c)^2 X^2)}{2 \pi^2 X^2} 
\Bigg\{
(1\!-\!c)^2
\left[
1\!-\!\,\frac{\beta_0 \alpha_s(X^2)}{2\pi}\right]^2
+O((1\!-\!c)^3)
\Bigg\}
\label{Kdip_Bal_expand_c1}
\, ,
\end{align}
keeping the logarithmic dependence in $(1\!-\!c)$.
Hence, one finds
\begin{align}
&{\cal K}^{B}(X,X){\cal K}^{B}(Y,Y)
-\left({\cal K}^{B}(X,Y)\right)^2
\nonumber\\
=&\,
{\cal K}^{B}(X,X){\cal K}^{B}(cX,cX)
-\frac{1}{4}\Big[{\cal K}^{B}(X,X)
+{\cal K}^{B}(cX,cX)
-K_{\rm dipole}^{\rm B} (X,cX)
\Big]^2
\nonumber\\
=&\,
-\frac{1}{4}\Big[{\cal K}^{B}(cX,cX)
\!-\!{\cal K}^{B}(X,X)\Big]^2
+{\cal K}^{B}(X,X)K_{\rm dipole}^{\rm B} (X,cX)
+O((1\!-\!c)^3)
\nonumber\\
=&\,
\frac{(1\!-\!c)^2\alpha_s(X^2)}{4 \pi^4 X^4} 
\Bigg\{
-\alpha_s(X^2) \left[
1\!-\!\frac{\beta_0 \alpha_s(X^2)}{4\pi}\right]^2
+\alpha_s((1\!-\!c)^2 X^2)
\left[
1\!-\!\frac{\beta_0 \alpha_s(X^2)}{2\pi}\right]^2
+O(1\!-\!c)
\Bigg\}
\nonumber\\
\sim &\,
- \frac{(1\!-\!c)^2\alpha_s(X^2)^2}{4 \pi^4 X^4} 
\left[
1\!-\!\frac{\beta_0 \alpha_s(X^2)}{4\pi}\right]^2
\leq 0
\label{nonlin_cond_rcBal_expand_c1}
\, .
\end{align}
Therefore, Balitsky's prescription for rcJIMWLK kernel \eqref{Eq:JIMWLKBal} violates the nonlinear condition \eqref{cond1}, and thus it is not positive semidefinite. 

In addition to the regime $c\rightarrow 1$ for $Y=c X$, one can consider for completeness also the regime $c\rightarrow 0$, corresponding to
$|z\!-\!y|\ll|x\!-\!z|\sim |x\!-\!y|$. In that regime, still assuming that $X$ is small enough for the one-loop expression \eqref{1loop_alpha_s} for the running coupling to apply, one finds the Taylor expansion
\begin{align}
\alpha_s((X-Y)^2) =  \alpha_s((1\!-\!c)^2 X^2)
= 
\alpha_s(X^2)\Bigg\{
1
&\,
-\frac{\beta_0 \alpha_s(X^2)}{2\pi}\,
c
+O(c^2)
\Bigg\}
\label{aXY_expand_c0}
\end{align}
for $c\rightarrow 0$.
By contrast, $\alpha_s(Y^2)$ goes to $0$ logarithmically in that limit, as
\begin{align}
\alpha_s(Y^2) =  \alpha_s(c^2 X^2)
\sim &\,
\frac{4\pi}{\beta_0}\,
\frac{1}{\ln\left(\frac{1}{c^2}\right)}
\rightarrow 0
\, .
\end{align}
Then, from Eq.~\eqref{Eq:JIMWLKBal}, one obtains 
\begin{align}
{\cal K}^{B}(X,Y)
=&\, {\cal K}^{B}(X,cX)
=\frac{\alpha_s(X^2)}{2 \pi^2 X^2}\, \frac{1}{c}\,
\left[1+\frac{\beta_0 \alpha_s(c^2X^2)}{4\pi}\right]
+O(1)
\label{K_Bal_XY_expand_c0}
\end{align}
by expanding in powers of $c$ for $c\rightarrow 0$, while keeping the full logarithmic dependence on $c$. 
Consequently,
\begin{align}
&
{\cal K}^{B}(X,X){\cal K}^{B}(cX,cX)
-\Big[{\cal K}^{B}(X,cX)
\Big]^2
\nonumber\\
=&\,
\frac{\alpha_s(X^2)}{4 \pi^4 X^4}\, \frac{1}{c^2} 
\Bigg\{
\alpha_s(c^2 X^2)
-\alpha_s(X^2) \left[
1\!+\!\frac{\beta_0 \alpha_s(c^2 X^2)}{4\pi}\right]^2
+O(c)
\Bigg\}
\nonumber\\
\sim &\,
- \frac{\alpha_s(X^2)^2}{4 \pi^4 X^4}\, 
\frac{1}{c^2}
< 0
\label{nonlin_cond_rcBal_expand_c0}
\, .
\end{align}
Hence, the nonlinear condition \eqref{cond1} is violated in both regimes $c\rightarrow 1$ and $c\rightarrow 0$, for $Y=c X$.

\subsection{NLO fixed-order JIMWLK kernel in the Kovchegov-Weigert scheme}
The Kovchegov-Weigert (KW) UV subtraction scheme~\cite{Kovchegov:2006vj} leads to the following NLO contribution to the ${\cal K}_{JSJ} (X,Y)$  kernel 
\begin{align}
        {\cal K}^{\rm NLO, KW}_{JSJ} (X,Y) =&\, 
        \frac{\alpha_s^2\beta_0}{8\pi^3}\, 
        \frac{1}{\left(X^2\!-\!Y^2\right)}        
        \left\{ 
\frac{X\!\cdot\! Y}{X^{2}Y^{2}}\,
\left[       
    X^2 \ln \left( X^2\hat{\mu}^2\right)
   - Y^2\ln \left(  Y^2\hat{\mu}^2\right)
\right]   
   -\ln
   \left(\frac{X^2}{Y^2}\right)
        \right\}
\nonumber\\
& 
+{\alpha_s^2\over 8\pi^3}\,
\frac{X\!\cdot\! Y}{X^{2}Y^{2}}\,
\left\{ \left(\frac{67}{9}\!-\!\frac{\pi^2}{3}\right)N_c - \frac{10}{9}\, N_f\right\}       
        \, .
\label{Eq:JIMWLK_KW_fixed_order_NLO}        
\end{align}
As discussed in section \ref{sec:scheme_dep}, UV divergences appearing at NLO in the JIMWLK Hamiltonian can and should be transferred to the kernel ${\cal K}_{JSJ} (X,Y)$ multiplying the color operator already present at LO. But this procedure is scheme dependent, so that ${\cal K}_{JSJ} (X,Y)$ becomes scheme dependent at $\alpha_s^2$ order, which explains the difference between the expression \eqref{Eq:JIMWLK_KW_fixed_order_NLO}
obtained within the Kovchegov-Weigert scheme for UV subtraction and the expression \eqref{Eq:JIMWLKBal_fixed_order_NLO}
obtained within the Balitsky scheme.
Note that the overall contribution proportional to $\beta_0 \ln{\hat{\mu}^2}$ is the same in both cases. Indeed, it represents the leftover from the UV divergence after the standard UV renormalization (in the $\overline{\rm MS}$ scheme). The scheme dependence is associated with finite contributions which might be subtracted and transferred to ${\cal K}_{JSJ} (X,Y)$ or not, together with the UV divergences.   For that reason, the contributions in $\beta_0 \ln{X^2}$ or $\beta_0 \ln{Y^2}$ differ between Eqs.~\eqref{Eq:JIMWLKBal_fixed_order_NLO} 
and 
\eqref{Eq:JIMWLK_KW_fixed_order_NLO}.
By contrast, the contribution proportional to the LO kernel, on the second line of Eqs.~\eqref{Eq:JIMWLKBal_fixed_order_NLO} 
and 
\eqref{Eq:JIMWLK_KW_fixed_order_NLO} is the same in both of these UV subtraction schemes.

Including the LO contribution \eqref{Eq:KLO} and the NLO correction \eqref{Eq:JIMWLK_KW_fixed_order_NLO}, one finds for the ${\cal K}_{JSJ} $ kernel in the Kovchegov-Weigert scheme, in the collinear configuration  $Y=c X$,
\begin{align}%
    {\cal K}^{\rm KW}_{JSJ}(X,X) {\cal K}^{\rm KW}_{JSJ}(c X,c X) - \left[ {\cal K}^{\rm KW}_{JSJ}(X,c X) \right]^2
    &=       \frac{\alpha_s^3\beta_0}{16\pi^5\, X^4}
    \frac{(1\!-\!c) \ln(c^2)
   }{c^2
   (1+c)}
  +O(\alpha_s^4) 
   \le 0 
   \label{violation_nonlincond_FO_KW}
\end{align}
due to Eq.~\eqref{Eq:ColinearLO}.
Indeed, the function of $c$ on the right hand side of Eq.~\eqref{violation_nonlincond_FO_KW} vanishes at $c=1$ and is strictly negative otherwise (and goes to $-\infty$ at $c\rightarrow 0$).
Hence, the fixed order NLO JIMWLK kernel in the KW UV subtraction scheme violates the nonlinear condition \eqref{cond1}, and thus it is not positive semidefinite.
Moreover, it also violates the condition \eqref{cond1bis}, in the same way as in Balitsky's UV subtraction scheme.

\subsection{Kovchegov-Weigert prescription for rcJIMWLK}

The Kovchegov-Weigert  prescription
for rcJIMWLK, based on a resummation of the NLO $\beta_0$ terms from Eq.~\eqref{Eq:JIMWLK_KW_fixed_order_NLO}, reads~\cite{Kovchegov:2006vj}
\begin{align}
    {\cal K}^{KW} (X,Y)= 
    \frac{1}{2\pi^2}
    \frac{\alpha_s(X^2) \alpha_s(Y^2)} {\alpha_s(R^2)}\frac{X\cdot Y} {X^2 Y^2}\,
    \label{K_KW}
\end{align}
where 
\begin{align}
    R^2 = |X|\, |Y|\, \left( \frac{Y^2}{X^2}\right)^{\Theta/2}\,,~~~~~~~~
    \Theta = 
    \frac{X^2  + Y^2} {X^2-Y^2} 
    - 2 \frac{X^2 Y^2}{X\cdot Y} \frac{1} {X^2-Y^2}\,.
\end{align}
Eq.~\eqref{K_KW}
implies that 
\begin{align}
    {\cal K}^{KW} (X,X)= 
    \frac{\alpha_s(X^2)}{2\pi^2\, X^2}
    \label{K_KW_XX}
    \, ,
\end{align}
which is the same expression as in the case of Balitsky's prescription for rcJIMWLK, see Eq.~\eqref{Eq:JIMWLKBal_XX}. Hence, the condition \eqref{cond1bis} is satisfied provided the coupling $\alpha_s(X^2)$ is prevented to go negative in the IR, with an appropriate improvement of the one-loop expression \eqref{1loop_alpha_s}.

Let us now check the validity of the nonlinear condition \eqref{cond1} for positive semidefiniteness, in the case of Eq.~\eqref{K_KW}, focusing on the configurations $Y=cX$. In that case, the scale $R^2$ reduces to $c^{2/(1+c)}  X^2$. Thus we have 
\begin{align}
\label{Eq :KWpositivity_1}
    &{\cal K}^{KW} (X,X) {\cal K}^{KW} (Y,Y) - \left[ {\cal K}^{KW} (X,Y) \right]^2  
    \notag \\ 
    = &\,
    \frac{\alpha_s^2(X^2) \alpha_s^2(c^2 X^2)}{4\pi^4 c^2 X^4}   
    \left[ \frac{1}{\alpha_s(X^2) \alpha_s(c^2 X^2)}
    - \frac{1}{\alpha_s^2(c^{2/(1+c)}  X^2)}
    \right]\,.
\end{align}
Assuming that $X^2$, $c^2 X^2$ and $c^{2/(1+c)}  X^2$ correspond to small distances, in the perturbative domain, one can use the one-loop expression \eqref{1loop_alpha_s} of the running coupling in order to extract the $c$ dependence, and find
\begin{align}
\label{Eq :KWpositivity_2}
    &{\cal K}^{KW} (X,X) {\cal K}^{KW} (Y,Y) - \left[ {\cal K}^{KW} (X,Y) \right]^2  
    \notag \\ 
    = &\,
    -\frac{\alpha_s^2(X^2) }{4\pi^4 c^2 X^4} 
    \frac{1}{\left[\frac{4\pi}{\beta_0\alpha_s(X^2)} + \ln\left(\frac{1}{c^2}\right)\right]^2}
\left\{\frac{4\pi}{\beta_0\alpha_s(X^2)}\,
\frac{(1\!-\!c)}{(1+c)}\ln\left(\frac{1}{c^2}\right)
+\left[\frac{1}{(1+c)}\ln\left(\frac{1}{c^2}\right)\right]^2
\right\}
   \,.
\end{align}
Noticing that 
\begin{align}
\frac{(1\!-\!c)}{(1+c)}\ln\left(\frac{1}{c^2}\right)
\geq 0\, ,
\end{align}
with the equality only at $c=1$, one finds the the expression
\eqref{Eq :KWpositivity_2} vanishes at $c=1$, and is strictly negative for other real values of $c$ (and goes to $-\infty$ for $c\rightarrow 0$).
Therefore, the condition \eqref{cond1} is violated, meaning that the Kovchegov-Weigert prescription
for rcJIMWLK \eqref{K_KW} is not positive semidefinite.

\subsection{Other resummation schemes for rcJIMWLK}
We have shown so far the violation of positive semidefiniteness for two known prescriptions for rcJIMWLK.
We also tried several other resummations schemes, starting from the NLO JIMWLK kernel  Eq.~\eqref{Eq:JIMWLKBal_fixed_order_NLO} with Balitsky's UV subtraction scheme. 
We found all of them to violate the condition \eqref{cond1}, and thus to violate positive semidefiniteness. Here we only list the most compelling resummations we  considered: 
\begin{itemize}
    \item BLM-type~\cite{Brodsky:1982gc} prescription for rcJIMWLK:
 it amounts to determine the value of the renormalization scale
 $\mu_{\overline{\rm MS}}=\mu_{\rm BLM}$
 for which the $\beta_0$ part of the NLO contribution \eqref{Eq:JIMWLKBal_fixed_order_NLO} vanishes.
Then, the obtained rcJIMWLK is simply the LO kernel, but with the coupling at that scale, $\alpha_s(\mu_{\rm BLM}^2)$ or, using the position space notation \eqref{def_alpha_s_pos_space}, $\alpha_s(R_{\rm BLM}^2)$, where 
$\mu_{\rm BLM}^2 = 4e^{-2 \gamma_E}/R_{\rm BLM}^2$.
The scale $R_{\rm BLM}^2$ is thus determined from Eq.~\eqref{Eq:JIMWLKBal_fixed_order_NLO} as
\begin{equation}
\begin{split}
& 0= {\alpha_s^2\beta_0\over 16\pi^3}\left\{
2  \frac{X\cdot Y}{X^2 Y^2}
\ln\frac{(X-Y)^2}{R_{\rm BLM}^2}
+ \frac{1}{X^{2}} \ln \frac{Y^{2}}{(X-Y)^2} +\frac{1}{Y^{2}} \ln \frac{X^{2}}{(X-Y)^2}  \right\}
\label{Eq:JIMWLK_BLM_1}
\, ,
\end{split}
\end{equation}
so that 
\begin{equation}
\begin{split}
& R^{2}_{\rm BLM} = 
(X-Y)^2
 \left(\frac{(X-Y)^2}{Y^{2}}\right)^{-\frac{Y^{2}}{2  X\cdot Y}}
 \left(\frac{(X-Y)^2}{X^{2}}\right)^{-\frac{X^{2}}{2  X\cdot Y}}
\label{Eq:JIMWLK_BLM_5}
\, .
\end{split}
\end{equation}

As a remark, the same exercise can be performed at the level of the BK kernel instead of the JIMWLK kernel. The result is provided, for example, in Ref.~\cite{Beuf:2017bpd}. However, using the BLM procedure at the level of BK or at the level of JIMWLK lead to different running coupling prescriptions. Neither of them lead to a positive semidefinite rcJIMWLK kernel.

\item Within Balistky's scheme for UV subtraction, Eq.~\eqref{Eq:JIMWLKBal_fixed_order_NLO}, one can write a triumvirate expression for rcJIMWLK, following Kovchegov and Weigert. One finds in such a way
\begin{align}
    {\cal K}^{tr} = 
    \frac{1}{2\pi^2}
    \frac{\alpha_s(X^2) \alpha_s(Y^2)} {\alpha(R_{tr}^2)}\frac{X\cdot Y} {X^2 Y^2}
\end{align}
where 
\begin{align}
    R_{tr}^2 = \frac{X^2 Y^2} {(X-Y)^2}
    \left( \frac{(X-Y)^2}{Y^2} \right)^\frac{Y^2}{2 X\cdot Y}
        \left( \frac{(X-Y)^2}{X^2} \right)^\frac{X^2}{2 X\cdot Y}\,.
\end{align}

\end{itemize}

Moreover, some of the running coupling prescriptions used in the literature are simply guesses, not related to NLO calculations. As an example, the parent dipole prescription for the BK equation has been often used. In appendix \ref{sec:PD}, the corresponding prescription for rcJIMWLK is written down, and found to violate positive semidefiniteness. 

\section{Discussion}

\label{Sec:discussions}

Phenomenological studies at high energy require computations at NLO accuracy for precision and running coupling effects are known to be very important beyond strict LO accuracy. In this paper, we reviewed some of the running coupling prescriptions for BK-JIMWLK evolution. The main tool used in our analysis is the requirement of positive semidefiniteness of the Hamiltonian. This requirement is tightly related with the possibility of constructing a Langevin formulation of the evolution, i.e. such a construction is only possible if the Hamiltonian is positive semidefinite. 

There are several sources of scheme dependence in the construction of a running coupling prescription for JIMWLK. The most crucial one turns out to be the freedom in the choice of basis of color operators used to write the NLO JIMWLK Hamiltonian. The operators in that basis should contain suitable subtraction terms, so that all UV divergences appearing at NLO are transferred to the coefficient of the operator already appearing at LO. There is a freedom in that procedure, associated with the possibility of transferring finite contributions together with the UV divergences. For that reason, the NLO correction ${K}_{JSJ}^{\rm NLO}(x,y,z)$ to the coefficient of the LO operator is scheme dependent, which leads to an ambiguity in the large logs found at NLO to be resummed into the running coupling. A further important source of scheme dependence is the freedom in writing an all-order resummed expression for rcJIMWLK based on the knowledge of NLO corrections only.

We have considered the JIMWLK kernels including the fixed order NLO correction ${K}_{JSJ}^{\rm NLO}(x,y,z)$ found either in Balitsky's scheme \cite{Balitsky:2006wa} for UV subtraction (see Eq.\eqref{Eq:JIMWLKBal_fixed_order_NLO}) or in the Kovchegov-Weigert scheme \cite{Kovchegov:2006vj} (see Eq.\eqref{Eq:JIMWLK_KW_fixed_order_NLO}). In both of these cases, there is a rather trivial violation of positive semidefiniteness, due to the violation of the condition \eqref{cond1bis}. This can be interpreted as the inconsistency, beyond LO, of fixed order JIMWLK evolution with fixed coupling in QCD. Moreover, the nonlinear condition \eqref{cond1} is violated in both of these UV subtraction schemes, providing a further obstruction to positive semidefiniteness. 
Note however that we have not included in our study the other NLO corrections to the JIMWLK Hamiltonian, associated with new color operators. Indeed, it is unclear how to check the positive semidefiniteness property of the entire NLO JIMWLK Hamiltonian.

We have also considered various prescriptions for rcJIMWLK, corresponding to the resummation of the $\beta_0$ contribution in ${K}_{JSJ}^{\rm NLO}(x,y,z)$ found in either the Balitsky or  Kovchegov-Weigert scheme, including the running coupling prescriptions originally proposed in Refs.~\cite{Balitsky:2006wa} and \cite{Kovchegov:2006vj}. In all cases, we found the nonlinear condition \eqref{cond1} to be violated, and thus the corresponding rcJIMWLK kernel not to be positive semidefinite. 
There might exist a positive semidefinite rcJIMWLK kernel corresponding to a resummation based on Eq.~\eqref{Eq:JIMWLKBal_fixed_order_NLO} or \eqref{Eq:JIMWLK_KW_fixed_order_NLO}, that we have not thought of. 
However, we consider this possibility somewhat unlikely, because in all rcJIMWLK prescriptions we considered, the violation of the nonlinear condition \eqref{cond1} seems to follow from its violation at the unresummed level, in either of these two UV subtraction schemes. 

In a recent work \cite{KLS}, a new UV subtraction scheme was proposed for the  construction of the color operator basis to write the NLO JIMWLK Hamiltonian, as an alternative to the UV subtraction schemes of Balitsky and of Kovchegov-Weigert. In that new scheme, only contributions with small parton pairs unresolved by the target are subtracted from the new color operators appearing at NLO. In that sense, the new scheme can be considered more physical than the previous ones, and it should reduce the risk of oversubtraction. Interestingly, in the basis of color operators constructed in that new scheme, { a part of} the obtained $\beta_0$ contribution in ${K}_{JSJ}^{\rm NLO}(x,y,z)$ can be naturally resummed into the daughter dipole prescription, corresponding to the replacement written in Eq.~\eqref{daughter} in the LO JIMWLK kernel: 
{
\begin{eqnarray}
  {K}_{JSJ}^{\rm NLO}(x,y,z)&=&  {\cal K}^{\rm daughter}(x,y,z) 
  + \delta K
 \nonumber \\ 
  {\cal K}^{\rm daughter}(x,y,z)& =& \frac{\sqrt{\alpha_s(X^2)\alpha_s(Y^2)}
    }{2\pi^2} \frac{X \cdot Y}{X^2 Y^2} 
\end{eqnarray}
Here, $\delta K$ accounts for extra $\beta_0$-dependent NLO terms, which are to be resumed by the DGLAP evolution in the projectile \cite{KLS}.
}
The daughter dipole prescription {
${\cal K}^{\rm daughter}$} for rcJIMWLK is positive semidefinite. In particular, the nonlinear condition \eqref{cond1} is obeyed thanks to a trivial generalization of Eq.~\eqref{pos_semidef_LO}. The daughter dipole prescription was successfully implemented to perform numerical simulation of rcJIMWLK (in its Langevin form) \cite{Lappi:2011ju}.

At this stage, the daughter dipole prescription \eqref{daughter} for rcJIMWLK is thus the only known running coupling prescription which is both positive semidefinite (with a Langevin formulation) and obtained from NLO calculations. This also provides further motivation for 
the scheme constructed in Ref.~\cite{KLS}. 
Moreover, this study also seems to suggest that the apparent impossibility of finding a positive semidefinite rcJIMWLK based on the UV subtraction schemes of Balitsky or Kovchegov-Weigert might be the signal of a problem in these two schemes, for example of over subtraction.

{
In this paper,  we found many examples in which 
 the dipole kernel in rcBK
is positive 
(in the sense of Eq.~\eqref{Eq:cond3_dip}) and does not lead to any apparent problems, whereas the corresponding prescription for rcJIMWLK is not positive semidefinite.  One could be then puzzled about what might be }
 the consequences of the violation of positive semidefiniteness condition. 
From the 
rcJIMWLK perspective, this seems to be a major obstacle since positive semidefiniteness is mandatory not only for any reasonable evolution Hamiltonian but also for a stochastic 
formulation, i.e. numerical implementation. On the other hand,  rcBK (without further NLO corrections) does not display
any instability or nonphysical behavior and is broadly believed to be a reliable phenomenological tool. This might mean that the problems related to the violation of positive semidefiniteness are hidden in the {
higher  order (beyond dipole) correlators of Wilson lines in the Balitsky's hierarchy,  the
$N_c$ suppressed terms, which are neglected in the BK evolution. To prove this, this would require analyzing Balitsky hierarchy and it is not clear how one can do it consistently other than performing the analysis of the JIMWLK Hamiltonian as we did in the current paper. 
}

{
Finally, one could think of reducing the Hilbert space on which the JIMWLK Hamiltonian acts to gauge invariant objects only. On such a reduced Hilbert space,  the kernel $\cal K$ in the Hamiltonian could be replaced with its dipole version $K_{\rm dipole}$. Unfortunately, this is  neither helpful with a Langevin formulation  nor with positive semidefiniteness of the Hamiltonian. The latter property is still violated: while (\ref{Eq:cond3_dip}) is satisfied (\ref{cond1}) is badly broken.
}

\acknowledgments

We thank I.~Balitsky, A.~Dumitru, A.~Kovner, and  Yu.~Kovchegov for stimulating discussions. V.S. thanks V.~Kazakov for  illuminating discussions. 

This material is based
upon work supported by the U.S. Department of Energy, Office of Science, Office of Nuclear
Physics through the Contract No. DE-SC0020081 (V.S.) and the Saturated Glue (SURGE) Topical Collaboration (V.S.). M.L. and V.S. got support from the Binational Science Foundation under grant \#2022132.  M.L.  is supported by the Binational Science Foundation grant \#2021789 and by the ISF grant \#910/23. GB is supported in part by the National Science Centre (Poland) under the research grant no 2020/38/E/ST2/00122 (SONATA BIS 10).
This work has been performed in the framework
of the MSCA RISE 823947 ``Heavy
ion collisions: collectivity and precision in saturation physics'' (HIEIC).

We thank ExtreMe Matter Institute (EMMI), Physics Department of  Muenster University (A. Andronic), Ben-Gurion University of the Negev, National Centre for Nuclear Research, Institute for Nuclear Theory at  University of Washington, 
and ECT$^*$ for their support and hospitality during various stages of completing this project.

\appendix

\section{LO JIMWLK from bi-local noise}
\label{App:bitosi}
Lets return to Eq.~\eqref{Eq:UdelBi} 
where in order to reproduce LO JIMWLK kernel we consider an anzatz $\Psi_i(X,V) = \frac{X_i}{X^2} f(X,V)$~\footnote{This procedure is equivalent to finding eigenvectors and eigenstates of the kernel.}
\begin{align}
    \int_{v} f(X, V) f(Y,V) =  {\cal K} (X,Y) \frac{X^2 Y^2}{X\cdot Y}\,.
\end{align} 
The LO Kernel divided by the WW kernel is a constant  
$ {\cal K} (X,Y) \frac{X^2 Y^2}{X\cdot Y} = \frac{\alpha_{s}}{\pi^2} $. 
A particular solution of this equation is 
$f(X, V) = f_1  (V)$ where $f_1( V)$ is any normalizable function, with  
\begin{align}
    \int_{v} f_1^2( V)  =  \frac{\alpha_{s}}{\pi^2}\,. 
\end{align} 
Substituting this into Eq.~\eqref{Eq:UdelBi}  we obtain 
\begin{align*}
&\int D\zeta 
    e^{ \Delta \left(
    i \int_{x,y,z} \frac{X_i}{X^2} f_1(Y) q^a(x,z) \zeta_i^a(y,z) - \frac{1}{2} \int_{x,z}  \zeta_i^a(x,z)\zeta_i^a(x,z)
    \right)
    }\\ 
    &= \int D\zeta  D \theta D \xi e^{i \Delta \int_z \theta^a_i(z) \left( 
    \xi^a_i(z) - \int_x f_1(x-z) \zeta^a_i(x,z)
    \right) } 
     e^{ \Delta \left(
    i \int_{x,z} \frac{X_i}{X^2}  q^a(x,z) \xi_i^a(z) - \frac{1}{2} \int_{x,z}  \zeta_i^a(x,z)\zeta_i^a(x,z)
    \right)
    }
    \\ 
    &=  \int D \theta D \xi e^{i \Delta  \int_z \theta^a_i(z)  
    \xi^a_i(z) -\frac{\Delta}{2} \int_{x,z} \theta^2(z) f_1^2(x-z)
   } 
     e^{ i\Delta %
     \int_{x,z} \frac{X_i}{X^2} q^a(x,z) \xi_i^a(z) 
    } 
    \\ 
    &= \int  D \theta D \xi e^{i \Delta  \int_z \theta^a_i(z)  
    \xi^a_i(z) - \frac{\alpha_s \Delta}{2\pi^2}  \int_{z} \theta^2(z) 
   }  
     e^{i \Delta %
    \int_{x,z} \frac{X_i}{X^2} q^a(x,z) \xi_i^a(z)
    } 
         \\ 
    &=  \int  D \xi e^{\Delta \left( i \int_{x,z} \frac{X_i}{X^2} q^a(x,z) \xi_i^a(z) - \frac{\pi^2}{2\alpha}   \int_z  
    \xi^2(z) \right)
   }  
            \\ 
    &=  \int  D \xi e^{\Delta \left( i \frac{\sqrt{\alpha_s}}{\pi} \int_{x,z} \frac{X_i}{X^2} q^a(x,z) \xi_i^a(z) - \frac{1}{2}   \int_z  
    \xi^2(z) \right)
   }  
\end{align*}
where in the last line we performed the rescaling of the variable $\xi$. This reproduces Eq.~\eqref{Eq:UHST}.

\section{Parent dipole prescription
\label{sec:PD}}
The parent dipole prescription (note that it does not reproduce NLO JIMWKL if expanded to the relevant order) reads 
\begin{align}
K_{\rm dipole}^{\rm PD} (x,y,z) 
&= 
\frac{\alpha_s((X-Y)^2)}{2\pi^2} \frac{(X-Y)^2} {X^2 Y^2}
\label{PD_rcBK}
\end{align}
for the BK kernel.
The corresponding rcJIMWLK kernel can be fully fixed by  Eq.~\eqref{dipoleinvert} and the condition that going to fixed coupling one reproduces LO JIMWLK kernel
\begin{align}
{\cal K}^{\rm PD} (x,y,z) 
&= 
-\frac12 K_{\rm dipole}^{\rm PD} (x,y,z) + \frac{\alpha_s(X^2)}{4 \pi^2} \frac{1}{X^2} +  \frac{\alpha_s(Y^2)}{4 \pi^2} \frac{1}{Y^2} \,.
\label{PD_rcJIMWLK}
\end{align}

In order to  check for the validity of the non-linear condition \eqref{cond1}, let us focus on the configuration $Y=c X$, with $c \to 0$.
Then, the BK kernel \eqref{PD_rcBK} reduces to
\begin{align}
K_{\rm dipole}^{\rm PD} (X,Y) 
&= 
\frac{\alpha_s((1-c)^2X^2)}{2\pi^2} \frac{(1-c)^2} {c^2 X^2}
=
\frac{\alpha_s(X^2)}{2\pi^2} \frac{1} {c^2 X^2} +O\left(\frac{1}{c}\right)
\, ,
\label{PD_rcBK_1}
\end{align}
whereas for the rcJIMWLK kernel one has
\begin{align}
{\cal K}^{\rm PD}(X,X) 
&= 
\frac{\alpha_s(X^2)}{2\pi^2} \frac{1} {X^2}
\\
{\cal K}^{\rm PD}(cX,cX) 
&=
\frac{\alpha_s(c^2X^2)}{2\pi^2} \frac{1} {c^2X^2}
\, ,
\label{PD_rcJIMWLK_1}
\end{align}
so that
\begin{align}
{\cal K}^{\rm PD}(X,cX) 
&= 
\frac{1}{4\pi^2 X^2} \frac{1} {c^2}
\Big[\alpha_s(c^2X^2)-\alpha_s(X^2)\Big]
+O\left(\frac{1}{c}\right)
\, .
\label{PD_rcJIMWLK_2}
\end{align}
Hence, for $c\rightarrow 0$, one finds
\begin{align}
&{\cal K}^{\rm PD}(X,X) 
{\cal K}^{\rm PD}(cX,cX)
-\Big({\cal K}^{\rm PD}(X,cX)\Big)^2
\nonumber\\
=&\,
-\frac{1}{16\pi^4 X^4} \frac{1} {c^4}
\Big[\alpha_s(c^2X^2)-\alpha_s(X^2)\Big]^2
+O\left(\frac{1}{c^3}\right)
\leq 0
\, .
\end{align}
Hence, the condition \eqref{cond1} is violated, meaning that the rcJIMWLK kernel \eqref{PD_rcJIMWLK} corresponding to the parent dipole prescription is not positive semidefinite.

\bibliography{Langevin}

\end{document}